\documentclass[12pt]{article}
\usepackage{amssymb}
\usepackage{amsmath}
\usepackage[dvips]{graphicx}
\begin{document}

\title{\LARGE\bf Broadening of band-gap in photonic crystals with optically saturated media}

\author{
\normalsize\bf S. M. Abrarov$^1$ and R. M. Abrarov$^2$}

\maketitle

\begin{abstract}
Due to strong absorption of the incident light, the media with high refractive index are considered restrictive for applications in photonic crystals (PhCs). The possibility to resolve this problem by optical saturation effectively minimizing the absorption of the PhC medium is discussed. Such approach might be promising for the significant broadening of the photonic band-gap.
\end{abstract}

\section {Introduction}
Many recent developments in the modern optics are closely related to the specific types of the metamaterials, called photonic crystals (PhCs). The existence of photonic band-gap (PBG), predicted by E. Yablonovitch \cite{Yablonovitch87} and S. John \cite{John87}, opens new opportunities for the practical applications of PhCs leading to the significant improvement of the efficiency and durability of the various opto-electronic devises.

The refractive index contrast (RIC), the ratio between the refractive indices of the component materials, is the most important parameter of PhCs, which increase results to the broadening of PBG as well as the solid angle of the total reflection \cite{Biswas97, Gaillot05}. Hypothetically, the use of argon instead of air as an ambient filling the voids of PhC can increase RIC since the refractive index of this gas can be less than unity in the spectrum of anomalous dispersion \cite{Guenther90}. Alternatively, some non-linear effects leading to the self-focusing of the light in the media with positive intensity-dependent refractive index \cite{Chiao64, Thompson72} such as optical fibers and semiconductor lasers can also be applied to increase RIC. However, none of these methods are efficient for the significant broadening of PBG due to their small contribution.

The use of the semiconductors with high refractive index is very restrictive due to rapid growth of the absorption coefficient at the resonance frequencies near the band-edge. In this work the possibility to overcome this problem by optical saturation of PhC medium is discussed.

It should be noted that optically saturated materials are widely used in the mode-locking devises in which the short-term pulses are generated whenever a saturable absorber is bleached out by the incident laser light of the high intensity \cite{Silfast04}.

\section {Mechanism of optical saturation}

\subsection{Discrete spectra}

The spectra of various optical media (e.g. CO$_2$, He-Ne, ruby, Nd-YAG) contain the set of the discrete energy levels. Therefore their absorption and emission properties are determined by optical transitions of the electrons originating between these discrete energy levels. 

Consider a simplest case, an atomic system consisting of two discrete levels, non-degenerate energy states $E_{1} $ and $E_{2} $, where $E_{2} >E_{1} $. Let us denote $N_{1} $ and $N_{2} $ as the number of electrons per unit volume occupying the states 1 and 2, respectively, and $N=N_{1} +N_{2} $ as the total number of electrons per unit volume. 

The equation describing the intensity of a plane electromagnetic wave propagating through a medium along \textit{z-}direction is given by
\[\frac{\partial }{\partial z} I=\gamma \left(\nu \right)I, \]
where $I$ is the intensity and $\gamma \left(\nu \right)$ is the gain. The solution of the equation above is
\begin{equation} \label{Eq1} I\left(z\right)=I\left(0\right)e^{\gamma \left(\nu \right)\, z},\end{equation}
where $I\left(0\right)$ is the intensity of the incident light. 

Applying Einstein A and B coefficients for spontaneous and stimulated emissions, the gain for the media with homogeneous broadening can be expressed as \cite{Silfast04, Yariv91}
\begin{equation} \label{Eq2}\begin{array}{rcl} {\gamma \left(\nu \right)} & {=} & {\left(N_{2} -N_{1} \right)\frac{c^{2} }{8\pi \nu ^{2} n^{2} t_{sp} } g_{L} \left(\nu \right)} \\ {} & {=} & {\frac{\gamma _{0} \left(\nu \right)}{1+I/I_{s} \left(\nu \right)} } \end{array} \end{equation}
where $c$ is the light velocity, $n$ is the refractive index, $t_{sp} $ is the spontaneous emission rate per unit volume, $g_{L} \left(\nu \right)$ is the lineshape function (Lorentz profile),$\gamma _{0} \left(\nu \right)$ is unsaturated gain, and $I_{s} \left(\nu \right)$ is the saturation intensity. 

Assuming that the incident light intensity is weak, the relation between \eqref{Eq1} and Beer's law \cite{Silfast04}
\[I\left(z\right)=I\left(0\right)e^{-\alpha z} ,\] 
where $\alpha $ is the absorption coefficient, can be readily established. In particular, if $I<<I_{s} $ the gain $\gamma \left(\nu \right)$ is independent of $I$ since $N_{1} >>N_{2} $ or $N_{2} -N_{1} \approx -N$. This leads to
\[\gamma \left(\nu \right)\approx -\alpha \left(\nu \right)=-N\frac{c^{2} }{8\pi \nu ^{2} n^{2} t_{sp} } g\left(\nu \right),\] 
in accordance with \eqref{Eq2}.

In two-level system the population inversion cannot be created: $N_{2} $ tends to but never exceeds $N_{1} $ with increasing the incident light intensity. Therefore in such a system the gain $\gamma \left(\nu \right)$ is always negative and the stimulated absorption prevails over the stimulated emission. However, if the input intensity $I$ is large enough and considerably greater than $I_{s} $, then the population of the upper level 2 is almost same as that of the level 1. On the other hand, if $N_{2} \approx N_{1} $, then $\gamma \left(\nu \right)\approx 0$ according to \eqref{Eq2}. This means that if the condition $I>>I_{s} $ is satisfied, then the photon absorption and emission rates are almost equal to each other signifying that the medium is bleached out, i.e. becomes practically transparent for the incident light.

The gain $\gamma \left(\nu \right)$ can be positive in the atomic system with three and more discrete energy levels in which the state of the population inversion is feasible. The media with population inversion are used in the traveling wave amplifiers, based on the laser diodes or active optical fibers.

\subsection{Continuous spectra}

The absorption and emission in the semiconductor media mostly originate through radiative transitions involving the valance and conduction bands, which spectra are continuous. The rigorous analysis of the quantum - mechanical processes of the absorption and emission, based on Schr\"odinger equation with Hamiltonian describing the interaction between carriers and electromagnetic field, is quite complex. Upon some approximations, however, the similar essential results, necessary for our discussion, can be obtained by using a semi-classical approach \cite{Yariv91}.

Suppose that the excessive carriers in a semiconductor, electrons and holes, are generated either by optical pump or current injection. The gain for the semiconductor at a state of non-thermal equilibrium is given by \cite{Yariv91}
\begin{equation}\label{Eq3}{\gamma \left(\omega _{0} \right)=\int _{0}^{\infty }\left(\hbar \omega -E_{g} \right)^{1/2} \left(\frac{2m_{r} }{\hbar ^{2} } \right)^{1/2} \frac{m_{r} \lambda _{0}^{2}\left[f_{c} \left(\omega \right)-f_{v} \left(\omega \right)\right]}{\pi ^{2} \hbar 4n^{2} \tau _{rec} }  \, g_{L} \left(\omega \right)d\omega ,}\end{equation}
where $\hbar $ is Plank's constant, $E_{g} $ is the electronic bang-gap, $m_{r} $ is the reduced effective mass, $\lambda _{0}$ is the wavelength corresponding to the angular frequency $\omega _{0} $ of the optical transition, $\it n $ is the refractive index, $\tau _{rec} $ is the recombination lifetime of the carriers, $g_{L} \left(\omega \right)$ is the lineshape function
\[\, g_{L} \left(\omega \right)=\frac{\tau _{m} }{\pi \left[1+\left(\omega -\omega _{0} \right)^{2} \tau _{m}^{2} \right]} ,\] 
determined by the mean lifetime $\tau _{m} $ for coherent interaction of $\bf k$-electrons with a monochromatic field, $f_{c} \left(\omega \right)$ and $f_{v} \left(\omega \right)$ are quasi-Fermi levels for the conduction and valance bands, respectively.

N. G. Basov et al. \cite{Basov61}, M. G. Bernard and G. Duraffourg \cite{Bernard61} independently showed the possible conditions for the gain in the semiconductors. For the condition $f_{c} \left(\omega _{0} \right)<f_{v} \left(\omega _{0} \right)$ the gain is negative: the absorption prevails over emission resulting to the decrease of the incident light intensity. For the condition $f_{c} \left(\omega _{0} \right)=f_{v} \left(\omega _{0} \right)$ the gain is zero: the incident light intensity remains unchanged with increasing $z$, i.e. the semiconductor is absolutely transparent for the propagating light. Finally, for the condition $f_{c} \left(\omega _{0} \right)>f_{v} \left(\omega _{0} \right)$ the gain is positive indicating the state of the population inversion: emission prevails over absorption resulting to the light amplification.

The equation \eqref{Eq3} can be approximated as \cite{Yariv91}
\[\gamma \left(\omega _{0} \right)=-\frac{k}{n^{2} } \chi ''\left(\omega _{0} \right),\]
where $k=2\pi n /\lambda _{0}$, and $\chi ''\left(\omega _{0} \right)$ is the imaginary part of the electronic susceptibility $\chi \left(\omega _{0} \right)=\chi '\left(\omega _{0} \right)-i\chi ''\left(\omega _{0} \right)$. The plane electromagnetic wave passing through a semiconductor with excessive carriers oscillates in the form $\sim \exp \left[i\left(\omega _{0} t-k'z\right)\right]$, where $k'$ is the complex wave number, represented in terms of the real and imaginary parts of electronic susceptibility
\[k'=k\left[1+\frac{\chi '\left(\omega _{0} \right)}{2n^{2} } \right]-i\frac{k\chi ''\left(\omega _{0} \right)}{2n^{2} } .\] 
The real part of the electronic susceptibility determines the phase of electromagnetic wave, while its imaginary part -- the change of the wave amplitude. 

It was shown theoretically that even a moderate absorption creates undesirable conditions for increase of PBG \cite{Tip00}. When the semiconductor is saturated, the complex part of the electronic susceptibility disappears resulting in a purely real value of the complex wave number. Hence the elimination of the complex part of the electronic susceptibility becomes favorable for PhC performance. 

If the population inversion is present in a medium, then the imaginary part of the electronic susceptibility $\chi ''\left(\omega _{0} \right)$ is negative leading to amplification of the incident light according to \eqref{Eq3}. In our best knowledge, the theoretical studies of the PhC medium with negative electronic susceptibility have never been reported yet. Therefore the behavior of the PBG at the presence of population inversion is unclear at the moment.

\section{Applications of saturated media}

There were several theoretical works predicting that for the optical spectra RIC above $\sim 3$ might be sufficient to observe the complete PBG in fcc-packed opal matrices \cite{Biswas97, Gaillot05, John88, Leung90, John99}. However, despite of the availability of such materials, the strong absorption of the incident light restricts their practical applications. Unfortunately, the increase of the refractive index is accompanied with the rapid growth of the absorption coefficient at the resonance frequencies near the band-edge of the semiconductor. Even though the penetration depth of the light matching PBG is very small (less than 100 wavelengths), the high value of the absorption coefficient considerably decreases the magnitude of the reflected light. For instance, the absorption coefficient of ZnO near the band-edge at 3.4 eV is $\sim 10^{5}$ cm $^{-1}$.   Thus even 100 nm ZnO film absorbs more than 50\% of the incident light energy. Therefore such materials cannot be applied for PBG broadening in a conventional way.

Presumably, this problem can be resolved by optical saturation of the PhC medium. Assume that the incident light at an angular frequency corresponding to the existing energy level within electronic band-gap $\omega _{0} <E_{g} /\hbar $ or above it $\omega _{0} \ge E_{g} /\hbar $ passes through the semiconductor. If the light intensity is high and significantly stronger than the saturation intensity, then, by analogy with two-level atomic system above, the difference between absorption and emission rates becomes negligible: the gain is very close to zero and the optically saturated semiconductor becomes practically transparent for the incident light.

The incident light matching PBG cannot penetrate deeper than the limited number of the layers in the periodical structure. As a result only the surface layers of PhC can be optically saturated. However most of the reflection takes place within the depth of these saturated layers. Consequently the reflected light matching PBG may have almost the same amplitude as that of the incident light.

Alternatively, the carriers can be excited externally either by optical pump at excitation angular frequency greater than that of the incident light $\omega _{exc} > \omega _{0} $ \cite{Comment} or by current injection. Similar to the previous case, the incident light can be in resonance with any existing energy level within the electronic band-gap $\omega _{0} <E_{g} /\hbar $ or above it $\omega _{0} \ge E_{g} /\hbar $.

The condition $f_{c} \left(\omega _{0} \right)>f_{v} \left(\omega _{0} \right)$ should have no any advantage over $f_{c} \left(\omega _{0} \right)\\=f_{v} \left(\omega _{0} \right)$. In fact, the gain coefficient practically feasible for the semiconductors is of the order of $\sim 10^{2}$ cm$^{-1}$. Since the penetration depth of the light matching PhC is very small, this value is insufficient for the observation of the reflected light amplification.

\section{Perspective materials}

There are semiconductors and their alloys (GaAs, GaP, GaSb, InP, InAs, InSb, ZnO, Al$_x$Ga$_{1-x}$As, In$_{1-x}$Ga$_x$As$_y$P$_{1-y}$, etc.) with sufficiently high refractive indices. Two of them, zinc oxide (ZnO) and gallium phosphide (GaP), have attracted recently a particular attention for applications in opal PhC structures.

Nowadays ZnO has become one of the most popular materials, applied in PhCs due to its remarkable optical properties. Specifically, ZnO is a unique semiconductor since it is very sensitive to the growing and annealing conditions whereby its spontaneous emission can be controlled. Depending on the degree of the crystallinity, ZnO predominantly emits in the UV or green spectra. Typically, the spontaneous emission of the polycrystalline ZnO, obtained by solution methods, exhibits the strong deep level emission in the green spectral region due to the high density of the crystal defects. In contrast, the single crystal provides dominant PL in the near UV region.

Concerning the earliest publications, the photoluminescence (PL) spectra of 2D periodical array of ZnO nano-wires, synthesized by immersing the porous alumina membrane, were shown in \cite{Shi00}. The similar configuration of ZnO nano-wires exhibiting the dominant ultraviolet (UV) PL was considered in \cite{Yuldashev02}. The structural properties of macroporous 2D ZnO films, electrodeposited by using polystyrene colloidal crystals as a template, were reported in \cite{Sumida01}. The novel chemical route for synthesis of ZnO sub-micron balls and their self-sedimentation for 2D and 3D opal films were described in \cite{Seelig03}. However the influence of PBG on the spontaneous emission of ZnO was left uncovered in these publications.

The growth procedure of the polycrystalline nanoparticles of ZnO in 3D bulk opal and the effect of PBG on the green \cite{Abrarov04} and UV \cite{Abrarov06} PL spectra were shown first in our publications. For the soaking of the samples we applied de-ionized water solution containing zinc nitrate hexahydrate (Zn(NO$_3$)$_2\cdot$6H$_2$O) precursor, used earlier by our research group \cite{Yuldashev02}. This technological process is simple and inexpensive but requires many cycles of the soaking and thermal annealing since the single one results in small, tenths of the percent, filling factor \cite{Feoktistov01}. Additionally due to congestion of the pores on the sample surface, the uniformity of ZnO deteriorates as infiltration increases. The CVD method, used for the growth of inverse ZnO opal \cite{Juarez05}, may be more effective for the uniform and dense deposition.

ZnO might be a promising candidate for observation of complete PBG because of its high refractive index exceeding 9 in the near UV spectral region. The first maximum of ZnO refractive index is observed at 367 nm, which is very close to the visible range \cite{Park68}. In order to shift the refractive index peak to the visible spectrum the alloy Zn$_{1-x}$Cd$_x$O can be employed. Due to close position of this peak to the edge of the visible spectrum (380 nm), even a small addition of cadmium may be sufficient to shift it towards the blue region. The decrease of Zn$_{1-x}$Cd$_x$O band-gap with increasing cadmium proportion can be estimated according to Vegard's law.

Another attractive semiconductor for applications in PhCs is GaP. In the near UV spectral region its refractive index exceeds 5 while over the most visible range it is around 3.5 \cite{Blistanov82}. Moreover, in the far infrared range near 27 $\mu$m, the refractive index of GaP grows very rapidly reaching a colossal value, about 16 at the maximum \cite{Guenther90, Blistanov82}. Therefore PhC based on GaP might be useful for the applications in microwave electronics. Recent publications show that MOCVD and atomic layer deposition methods provide good quality and uniform distribution of GaP inside the voids of opal matrices \cite{Yates05, Palacios05, Graugnard}.

\section{Conclusion}

A strong absorption of an incident light at the resonance frequencies near the semiconductor band-edge restricts the applications of the media with high refractive index in PhCs. The absorption of the incident light might be minimized by optical saturation of PhC medium either by direct exposure of the incident light with extremely high intensity or by external pump. The optical saturation might be promising for the significant broadening of PBG.

\bigskip

\noindent --------------------------------------------------------------------\\
$^1$ S. M. Abrarov, York University, Toronto, Canada\\
\indent\small Email: abrarov@yorku.ca\\
\\
$^2$ R. M. Abrarov, University of Toronto, Canada\\
\indent\small Email: rabrarov@physics.utoronto.ca


\begin{thebibliography}{26}
\bibitem{Comment}
Excited photon energy can be above PBG
\end{thebibliography}
\end{document}